# Remarkably strong magnetic response in molecules with polar groups

Jihong Wang[1†], Yizhou Yang[1†], Jie Jiang[2†], Liuhua Mu[3], Guosheng Shi[4], Yongshun Song[1], Haijun Yang[5], Peng Xiu[6], Liang Chen[2*], and Haiping Fang[1,7*]

[1]*School of Physics and School of Materials Science and Engineering, East China University of Science and Technology, Shanghai 200237, China*
[2]*School of Physical Science and Technology, Ningbo University, Ningbo 315211, China*
[3]*CAS Key Laboratory of Interfacial Physics and Technology, Shanghai Institute of Applied Physics, Chinese Academy of Sciences, Shanghai 201800, China*
[4]*Shanghai Applied Radiation Institute, Shanghai University, Shanghai 200444, China*
[5]*Interdisciplinary Research Center, Shanghai Synchrotron Radiation Facility, Zhangjiang Laboratory (SSRF, ZJLab), Shanghai Advanced Research Institute, Chinese Academy of Sciences, Shanghai 201204, China*
[6]*Department of Engineering Mechanics, Zhejiang University, Hangzhou 310027, China*
[7]*Oujiang Laboratory and Wenzhou Institute, University of Chinese Academy of Sciences, Wenzhou, Zhejiang 325000, China*

[†]These authors contributed equally to this work.
[*]Corresponding authors. Email: fanghaiping@sinap.ac.cn; liangchen@zafu.edu.cn

**For more than a century, electricity and magnetism have been believed to always exhibit inextricable link due to the symmetry in electromagnetism[1]. At the interface, polar groups that have polar charges[2], are indispensable to be considered, which interact directly with other polar charges/external charges/external electric fields. However, there is no report on the corresponding magnetic properties on these polar groups. Clearly, such asymmetry, that is, only the interaction between the polar groups and charges, is out of bounds. Here we show that those molecules with considerable polar groups, such as cellulose acetate (CA) and other cellulose derivatives with different polar groups, can have strong magnetic response, indicating that they are strongly paramagnetic. Density functional theory (DFT) calculation shows that the polarity greatly reduces the excitation energy from the state without net spin (singlet) to the state with net spin (triplet), making the considerable existence of magnetic moments on the polar groups. We note that the hydrophobic groups in these molecules have no magnetic moments, however, they make the molecules aggregate to amply the magnetic effect of the magnetic moments in the polar groups, so that these magnetic**

**moments can induce the strong paramagnetism. Our observations suggest a recovery of the symmetry with inextricable link between the electricity and magnetism at the interface. The findings leave many imaginations of the role of the magnetic interaction in biological systems as well as other magnetic applications considering that many of those polar materials are biological materials[2,3], pharmaceutical materials[4], chemical raw materials[5], and even an essential hormone in agricultural production[6].**

Polar groups are a primary building block of nature on the molecular scale, studied extensively in biological systems, biomedicine, chemical industry, physical science and materials science[2,3]. The interactions between the polar charges in polar groups and other polar charges/external charges/external electric fields are the main components of interactions in various materials. This is particularly fundamental for the biological system, as these interactions related to polar charges are the typical non-covalent interactions and the non-covalent interactions play a major role in dictating the advanced structures and functions of biological systems[3]. However, there is not any corresponding magnetic interaction associated with those polar groups reported. This state of affairs is clearly counterintuitive, because electricity and magnetism are inextricably linked together due to the symmetry in electromagnetism for more than a century[1].

Here we show that molecules with polar groups, including cellulose acetate (CA, $[C_6H_7O_2(OCOCH_3)_x(OH)_{3-x}]_n$), cellulose acetate butyrate (CAB, $[C_6H_7O_2(OCOCH_3)_x(OCOC_3H_7)_y(OH)_{3-x-y}]_n$), chitosan (CS, $[C_6H_{11}NO_4]_n$), ethyl cellulose (EC, $[C_6H_7O_2(OC_2H_5)_x(OH)_{3-x}]_n$), p-ethylbenzoic acid (PEA, $C_9H_{10}O_2$), o-toluene acetic acid (OTAA, $C_9H_{10}O_2$), and auxin (IAA, $CH_9NO_2$), which have the polar groups such as acetyl, butyryl, amino, ethyl, formate, hydroxyl, and methyl groups, have strong magnetic response when they aggregate into micro- and nano-scale particles in aqueous solution. Density functional theory (DFT) calculation shows a considerable existence of magnetic moments on polar groups due to the polarity. Our observations suggest a recovery of the symmetry with inextricable link between the electricity and magnetism at the interface, and leave many imaginations of magnetic applications.

CA suspensions were prepared for the experiments. 1.7 g CA powders, which comprise of 39.8 wt% acetyl ($CH_3$-CO-) and 3.5 wt% hydroxyl (-OH), were firstly added into 50 ml ultrapure water and shaken for 1 min at 25 °C, followed by settling for 30 min, and removed particles floating on the water. Then, 10 ml mixture was taken from the bottom into a four-sided transparent quartz cuvette (135 mm × 135 mm × 14 mm) (see Fig. S1), and subsequently 150 ml ultrapure water was added for dilution, shaken well and settled for 0.5 min to obtain a CA suspension. Next, a rectangular neodymium magnet (50 mm × 50 mm × 30 mm, surface magnetic field ~0.5 T) was placed outside of the cuvette (Fig. 1a). We consistently observed that CA particles were significantly attracted towards the magnet and many CA particles accumulated on the wall of the cuvette close to the magnet. The accumulated CA particles formed a square shape consistent with the shape of the magnet within ~10 min. We then moved the magnet slowly to the right, and it can be seen clearly that many of the attracted CA particles moved following the magnet accordingly (Fig. 1c and Supplementary Video 1). To make the behavior more clear, we superimposed snapshots at different times, as shown in Fig. 1d. These results indicate that CA suspensions can be attracted, accumulated and moved accordingly along with an external magnetic field caused by the magnet, showing a remarkably strong magnetic attraction response.

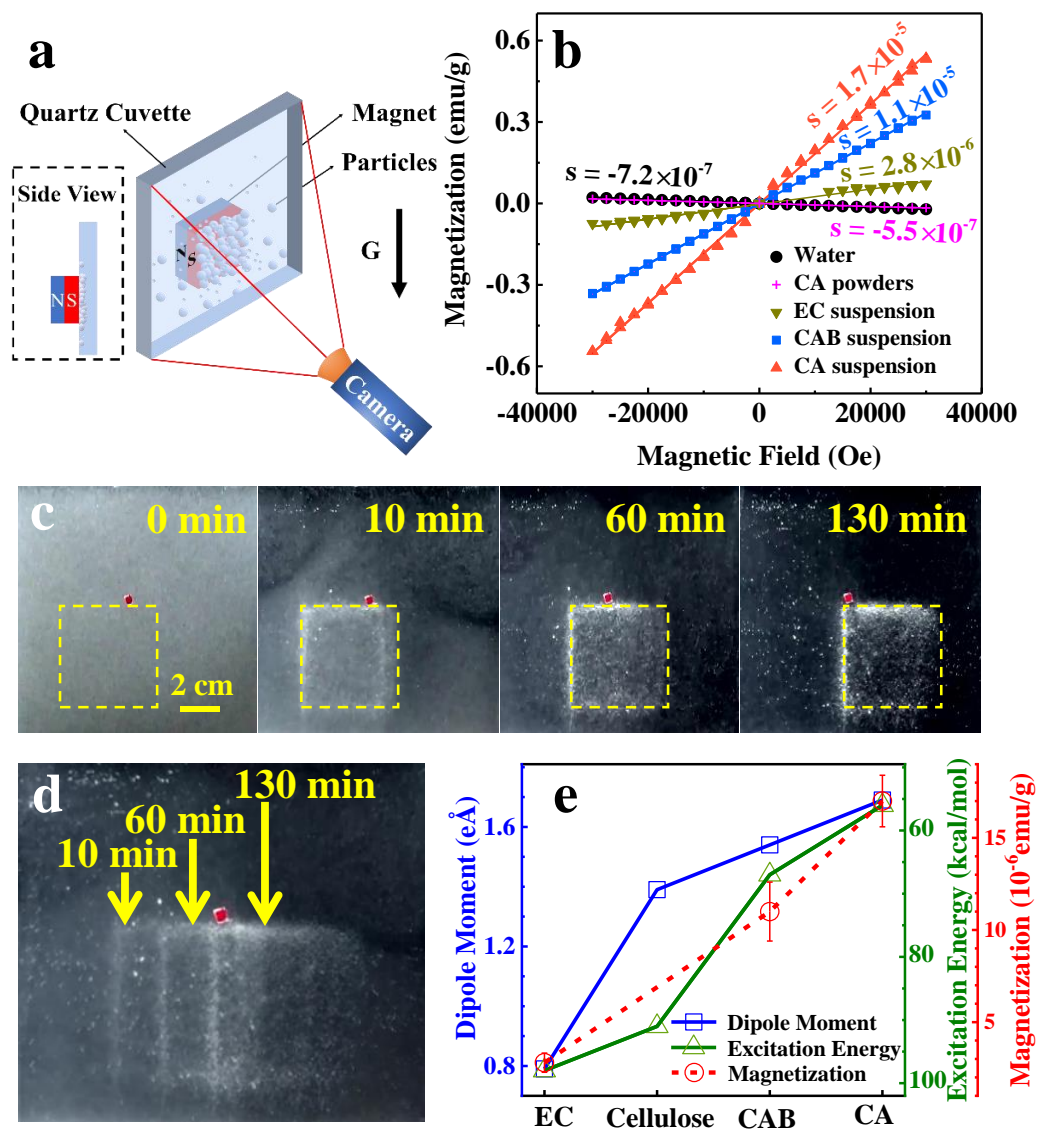

**Figure 1. Magnetic behavior of suspensions of cellulose acetate (CA) and other cellulose derivatives dispersed in ultrapure water. a,** Schematic picture of experimental set-up. The symbol 'G' represents the gravity. **b**, Magnetization ($M$) versus magnetic field ($H$) curves for cellulose acetate ($[C_6H_7O_2(OCOCH_3)_x(OH)_{3-x}]_n$), ethyl cellulose (EC, $[C_6H_7O_2(OC_2H_5)_x(OH)_{3-x}]_n$) and cellulose acetate butyrate (CAB, $[C_6H_7O_2(OCOCH_3)_x(OCOC_3H_7)_y(OH)_{3-x-y}]_n$) dispersed in ultrapure water or CA powder in a representative experiment, using water as a reference. The symbol '$s$'

represents the slope of the fitted lines, i.e. the mass susceptibility ($\chi_m$). **c,** Movement of CA suspension with the pull of magnet. The yellow dashed square indicates the position of the magnet, and the tiny white dots are CA particles. The small red square is a fixed position on the cuvette for guiding eyes. **d**, Superimposed snapshots of the CA suspension at different times from (c). **e**, Dipole moment, excitation energy from the singlet to triplet state, together with magnetic susceptibility of cellulose derivatives including EC, cellulose ($[C_6H_7O_2(OH)_3]_n$), CAB, and CA. The susceptibility of cellulose was not shown because we cannot obtain pure self-assembled cellulose particles in aqueous solution.

In order to quantitatively characterize the magnetism, we measured magnetization ($M$) versus magnetic field ($H$) curves by the superconducting quantum interference device (SQUID) magnetometer. Mass susceptibility ($\chi_m$) was then computed as $\chi_m = M/H = a/mH$, where $a$ is the moment measured by the SQUID magnetometer and $m$ is the mass of the CA powders (details in PS1 of Supplementary Information, SI). The magnetic susceptibility of the CA suspension exhibited strong paramagnetism of $(1.7 \pm 0.2) \times 10^{-5}$ emu/g, which was about 25 times larger than the absolute value of ultrapure water (Fig. 1b). In contrast, CA powders in dry state showed a diamagnetism with a magnetic susceptibility of $(-5.5 \pm 0.2) \times 10^{-7}$ emu/g. To exclude the possible contamination, we measured the concentrations of Fe, Co, and Ni in the CA powders using XPS (Fig. S2) and ICP-MS (Table S1), which were negligibly small.

The existence of such strong magnetic response of the CA suspension is surprising, but might not be so unexpected if we hope the electric and magnetic behavior are inextricably linked together. We all know that the spins of two electrons in the same occupied orbital must be antiparallel due to the Pauli exclusion principle, thus there is no net magnetic moment from the electron spin when the molecule is at singlet state with a total spin of $S = 0$. However, for the system with polar charges, electrons on atoms deviate from spatial symmetry, although the electrons are still paired. When the spatial symmetry is broken, we cannot still expect the two paired electrons consistent in a singlet state. Consequently, polar groups in those molecules, which contain polar charges, may have triplet states with a total spin of $S = 1$ that the spins of two electrons are parallel, which corresponds to the existence of magnetic moments, and we can expect that the stronger the polarity, the larger the possibility of the triplet states. Our DFT calculations demonstrated this by computing the excitation energy from the singlet to triplet state for cellulose derivatives with different polar substituents. As shown in Fig. 1e, the excitation energies were 97.6 kcal/mol, 91.2 kcal/mol, 66.8kcal/mol, and 56.4 kcal/mol for EC, cellulose, CAB, and CA, respectively (details in PS4 of SI). Considering that the larger the excitation energy, the smaller the possibility of triplet state (total spin of $S = 1$), the order from smallest to largest possibility of the triplet state was EC < cellulose < CAB < CA. Correspondingly, we calculated the dipole moments, which were 0.79 eÅ, 1.39 eÅ, 1.54 eÅ, and 1.69 eÅ for EC, cellulose, CAB, and CA (details in PS5 of SI), respectively. This shows a positive correlation between the polarity and the possibility of triplet state, as well as between the polarity and magnetic moment.

It should be noted that the excitation energies are still very high and the probabilities corresponding the energies are very small. Fortunately, the existence of the surrounding water will further reduce the barriers. Take the CA for example, with the help of the surrounding water, the excitation energy reduced from 56.4 kcal/mol to 13.3 kcal/mol for CA (details in PS6 of SI). This value was about two times the energy of hydrogen bond[7]. We note that the external magnetic field can also reduce the excitation energy[8,9]. Consequently, there is a considerable probability of triplet state as well as the magnetic moment. It is not easy and may be impossible to estimate the magnetic susceptibility of CA only from the magnetic moment, so we are still not sure whether the magnetic moments resulting from singlet-triplet transition are larger enough for the strong paramagnetism we observed experimentally and whether there are other mechanisms play key or dominated roles. Clearly, further studies are required.

The existence of strong magnetic response can be found in other two cellulose derivatives with polar substituents. Our experiments show that the suspensions of CAB and EC can also be attracted by a magnet (see Fig. S3) and the magnetic susceptibilities of them are $(1.1\pm0.2)\times10^{-5}$ emu/g and $(2.8\pm0.6)\times10^{-6}$ emu/g (see Fig. 1b), respectively. Interestingly, the magnetic susceptibility has the same order with the polarity of molecules as well as the excitation energy, as shown in Fig. 1e. We note that, we are not able to obtain a reliable value of the magnetic susceptibility of the cellulose, since the cellulose particles will sink to the bottom in solution that the aggregated cellulose suspension cannot be separated from the sunk cellulose particles.

In spite of the magnetic moment exists in the polar acetyl group, it is not certain that the existence of magnetic moments can lead to the observed strong magnetic response, as the interaction is very weak between a single magnetic moment and an external magnetic field, so that a single magnetic moment is very difficult to be detected at present instruments[10,11]. The hydrophobic groups have no magnetic moments; however, they make the CA molecules aggregate into micro- and nano-scale particles so that a large number of magnetic moments gather in a very small space, thus the response to the external field is greatly enhanced, like the superparamagnetic materials in the state of the nanoparticles[12]. Consequently, the CA suspensions exhibit strong magnetic response. Our experiments with a magnet (Fig. 1c and Fig. S3) show that the CA and other cellulose derivatives have aggregated into micro- and nano-scale particles in aqueous solution so that they can be directly observed by eyes. This is the reason why the magnetic response cannot be detected at present instruments for the acetic acid completely dissolved in water, since there are only fewer hydrophobic groups in the acetic acid molecules.

Thus, those molecules comprised of both polar groups and sufficient hydrophobic groups, potentially present a strong magnetic response in aqueous solution. The polar groups provide magnetic moments. The hydrophobic groups have no magnetic moments; however, they make the molecules aggregate into micro- and nano-scale particles to greatly amplify the interaction between magnetic moments and the external magnetic field. Based on this understanding, we have found many such molecules with considerable magnetic attraction responses including CS, PEA, OTAA, and IAA dispersed in ultrapure water (see Fig. S3), suggesting a universal magnetic response for molecules with polar groups in nature.

To summarize, we show that molecules with polar groups have strong magnetic response, indicating that in a polar group there is a considerable magnetic moment in addition to the polar charge. This suggests a recovery of the symmetry with inextricable link between the electricity and magnetism at the interface. However, different from the polar charge on a polar group which has a considerable interaction with external electric field, the magnetic moment of a polar group only has a very weak interaction with the external magnetic field and a nano- to micro-scale cluster is required to obtain a considerable response to the external magnetic field.

The magnetism for those molecules with polar groups, which we discovered here, is a long-overlooked interaction in biological system that plays a major role in dictating the advanced structures and functions as one of fundamental weak non-covalent interactions. Moreover, pharmaceutical materials and marine biological resources usually have polar groups such as $CS^{4,13}$. Therefore, we suggest that the strong magnetic responses of those materials can provide the new insights in explaining the magnetic effects of human body[14,15], promoting the magnetism related medicine, and designing advanced functional biomaterials. Besides, CA is the chemical raw material[5], and IAA is an essential hormone in agricultural production[6], the findings of their magnetic property may evoke novel magnetic applications, such as in the renewable energy development, utilization of marine and lake resources, wastewater purification and separation, and novel industrial technology. Clearly, our findings can be applied to a variety of fields, waiting for us to explore.